\documentclass[a4paper,  %
 reprint,
 amsmath,amssymb,
 aps, nofootinbib, twocolumn
]{revtex4-2}

\usepackage{graphicx}% Include figure files
\usepackage{dcolumn}% Align table columns on decimal point
\usepackage{bm}% bold math
\usepackage{xcolor}
\usepackage{amsmath}
\usepackage{comment}
\usepackage{enumitem}
\usepackage{float}
\usepackage{hyperref}
\usepackage{footnote}
\newcommand{\ed}{\end{document}}
\newcommand{\beq}{\begin{equation}}
\newcommand{\eeq}{\end{equation}}
\newcommand{\beqa}{\begin{eqnarray}}
\newcommand{\eeqa}{\end{eqnarray}}
\newcommand{\bc}{\begin{center}}
\newcommand{\ec}{\end{center}}

\newcommand{\ba}{\begin{array}}
\newcommand{\ea}{\end{array}}
\newcommand{\pa}{\partial}

\begin{document}

\preprint{APS/123-QED}

\title{ Spinning Black Hole in a Fluid  
}% Force line breaks with \\

\author{Surojit Dalui$^a$}
\email{surojitdalui@shu.edu.cn}
\author{Arpan Krishna Mitra$^b$}
\email{arpankmitra@aries.res.in}
\author{Deeshani Mitra$^c$}
\email{deeshani1997@gmail.com}
\author{Subir Ghosh$^c$}%
 \email{subirghosh20@gmail.com}
\affiliation{$^a$Department of Physics, Shanghai University, 99 Shangda Road, Baoshan District, Shanghai 200444, People Republic of China.}
\affiliation{$^b$Aryabhatta Research Institute of Observational Sciences (ARIES),\\ Manora Peak Nainital - 263001, Uttarakhand, India. }
\affiliation{$^c$Indian Statistical Institute, 203, Barrackpore Trunk Road, Kolkata 700108, India}

%\author{Deeshani Mitra}
%\email{deeshani1997@gmail.com}
%\affiliation{Indian Statistical Institute, \\ 203, Barrackpore Trunk Road, Kolkata 700108, India}
%\author{Surojit Dalui}
%\email{surojitdalui@shu.edu.cn}
%\affiliation{Department of Physics, Shanghai University, 99 Shangda Road, Baoshan District, Shanghai 200444, People Republic of China.}
%\author{Subir Ghosh}%
% \email{subirghosh20@gmail.com}
%\affiliation{Indian Statistical Institute, \\ 203, Barrackpore Trunk Road, Kolkata 700108, India}
%\author{Arpan Krishna Mitra}
%\email{arpankmitra@aries.res.in}
% \affiliation{Aryabhatta Research Institute of Observational Sciences (ARIES),\\ Manora Peak Nainital - 263001, Uttarakhand, India. }

%\author{Deeshani Mitra}
%\email{deeshani1997@gmail.com}
%\author{Surojit Dalui}
%\email{surojitdalui003@gmail.com}
%\author{Subir Ghosh}%
% \email{subirghosh20@gmail.com}
%\affiliation{Indian Statistical Institute, \\ 203, Barrackpore Trunk Road, Kolkata 700108, India}
%\author{Arpan Krishna Mitra}
%\email{arpankmitra@aries.res.in}
% \affiliation{Aryabhatta Research Institute of Observational Sciences (ARIES),\\ Manora Peak Nainital - 263001, Uttarakhand, India. }%Lines break automatically or can be forced with \\


\begin{abstract}
\noindent	
In this paper, we propose a new Analogue Gravity example - a spinning (or Kerr) Black Hole in an extended fluid model. The fluid model receives Berry curvature contributions   and applies to electron dynamics in Condensed Matter lattice systems in the hydrodynamic limit. We construct the acoustic metric for sonic fluctuations that obey a structurally relativistic wave equation in an effective curved background. In a novel approach of dimensional analysis, we have derived explicit expressions for effective mass and angular momentum per unit mass in the acoustic metric (in terms of fluid parameters), to identify with corresponding parameters of the Kerr metric. The spin is a manifestation of the Berry curvature-induced effective noncommutative structure in the fluid. Finally we put the Kerr Black Hole analogy in a robust setting by revealing explicitly the presence of horizon and ergo-region for a specific background fluid velocity profile. We also show that near horizon behavior of the phase-space trajectory of a probe particle agrees with  Kerr Black Hole analogy. In  fluid dynamics perspective, presence of a horizon signifies the wave blocking phenomenon.
\end{abstract}

\maketitle

%\tableofcontents

%\section{\label{sec:level1}Introduction:
%\protect\\  %\lowercase{via} 
\noindent
Analogue Gravity   \cite{rev} started with the work of Unruh \cite{unru},  who showed that first-order fluctuations in irrotational, non-viscous, barotropic flow obey a structurally relativistic massless scalar wave equation in an effectively curved background, with an Acoustic Metric (AM), comprising of fluid flow parameters (for diverse models, see \cite{analog,recent}). AM reveals  Black/White Hole - like features in velocity space, known as wave-blocking in fluid dynamics \cite{wb}.

In this paper, we construct a new AM, (in the framework of \cite{unru}),  in an extended fluid model with Berry curvature effects, derived in \cite{mitra2021}.  This phase space describes semi-classical electron dynamics in a magnetic Bloch band, with periodic potential in an external magnetic field and Berry curvature \cite{duval}.  This fluid dynamics is relevant in electron hydrodynamics in condensed matter,   where electron flow obeys hydrodynamic laws instead of  Ohmic  \cite{das}. Generically electrons in metals act as nearly-free Fermi gas with a large mean free path for electron-electron collision.  Recently hydrodynamic regime has been achieved in extremely pure,   high quality, electronic materials - especially graphene \cite{graf}, layered materials with very high electrical conductivity such as metallic delafossites  $PdCoO_2 ,~ PtCoO_2$ \cite{met}. 
 
The salient feature in our work is that the AM after a coordinate transformation \cite{Natario2009} is similar to Kerr metric \cite{kerr} in Eddington-Finkelstein (EF) coordinates \cite{Dalui2022}. Recently there have been several attempts to construct analogue models of BHs other than the non-rotating ones \cite{viss, rsc, swe, trs}. The fluid, in the presence of a vortex in it, has been considered as a system to construct an analogue of a rotating BH \cite{viss}. In \cite{rsc}, authors have reasoned, that in a shallow water system, with a varying background flow velocity, metric analogues of Kerr metric can be constructed. Later, the presence of superradiance has been found in it \cite{swe} and as well as in BEC \cite{trs}. In a  recent work in this direction but exploiting optical vortex is \cite{mor}, the authors have used Laguerre–Gaussian type beams, bearing phase singularities. These types of beams have transverse intensity profiles comprising all characteristics of a vortex.  The fluctuations in the amplitude and the phase of the electric field have been shown to satisfy a massless scalar field equation on a curved background, similar to the Kerr metric. However, this the present paper is possibly the first instance of an analogue Kerr metric in the fluid {\bf{subjected to an external magnetic field and Berry curvature.}} However, it is not unexpected since a spin-like feature appears in Berry curvature-modified particle dynamics \cite{sg}. The physics behind this AM is revealed 
through explicit construction of Kerr-like parameters, such as effective mass $m_{eff}$ and angular momentum per unit mass $a_{eff}$, out of fluid composites via dimensional analysis. More interestingly, using a specific form of non-uniform background fluid velocity we explicitly provide spatial positions of the ergo-region and horizon, characteristic of the Kerr metric. Recently, multiple articles have shown  studies on the trajectories of Weyl fermions in curved spacetimes.\cite{
volovik2016, Guan2017, Haller2023}. One of them has presented the trajectory of the massless Weyl particles around an analogue Schwarzschild bh \cite{Haller2023}. Here we have depicted the phase space trajectories of probe particle around the (Berry curvature induced) analogue Kerr metric that we have found and have pointed out that the location of the analogue horizon  in this analogue Kerr metric are same as that of one in Kerr metric in General Relativity.  \\
  {\it{AM with Berry curvature effects:}} We consider a  fluid with pressure $P(\rho)$. $e$ is electronic charge, ${\bf{B}}$  external magnetic field and  ${\bf{\Omega}}({\bf {k}})$ is  Berry curvature  in momentum $(\Bar{k})$ space.   For small ${\bf\Omega(k)} $  the extended fluid model  (with full expressions \cite{mitra2021} in Supplemental Material eqs.(1-3)) is,  $(\mathcal{A}({\bf x,\bf k})=1+e{\bf{B}}({\bf x})\cdot{\bf{\Omega}(\bf k)})$,
\begin{equation}
\label{bcont}
     \dot\rho=-\nabla\left(\frac{\rho \bf{v}}{\mathcal{A}}\right),
\end{equation}
\begin{equation}
\label{beul}
    \dot{{\bf{v}}}+ \frac{({\bf{v}}\cdot{\bf{\nabla}}){\bf{v}}}{\mathcal{A}}=-\frac{{\bf{\nabla}}P}{\rho\mathcal{A}}~. 
\end{equation}
     Irrotational  $\bf{v}=-\nabla \psi$ is written by a velocity potential $\psi$. Velocity $c_s$ of sonic disturbance in the medium and the system enthalpy $h$ are $ c_s=\sqrt{\frac{dP}{d\rho}}, ~\nabla h=\nabla P/\rho $
and \eqref{beul} becomes
\begin{equation}
\label{valpo2}
     -\nabla{\dot{\psi}}+\nabla\left[\frac{(\nabla{\psi})^{2}}{2\mathcal{A}}\right]=-\nabla{\frac{h}{\mathcal{A}}}   \rightarrow   \dot\psi-\frac{(\nabla{\psi})^{2}}{2\mathcal{A}}={\frac{h}{\mathcal{A}}}.
\end{equation}
 With fluid variables as  $ background~ +~  fluctuation $ \cite{unru}, 
\begin{eqnarray}
\nonumber
 \rho =\rho_0+\epsilon \rho_1, P=P_0+\epsilon c_s^2\rho_1, ~~~~~~~~~~~~~~~~\\
 v_{i}={v_{0i}}+\epsilon  v_{1i}=\partial_i\psi_0+\epsilon\partial_i\psi_1 , \nabla h_1=c_s^2\nabla \rho_1/\rho_0.
\end{eqnarray}

 First-order perturbation terms  are
 \begin{equation}
    {\rho}_{1}=\left(\frac{\rho_{0}\dot{\psi_{1}}}{c_{s}^{2}}\right)+\left(\frac{\rho_{0} \vec{v_{0}} \cdot \nabla{\psi_{1}}}{c_{s}^{2}\mathcal{A}}\right)~.
    \label{nabol}
\end{equation}
%$$\dot \rho_1 +\nabla.[\rho_0\{\nabla\psi /A +e(f.\nabla\psi )B\}+\rho_1\{v_0/A+e(f.v_0)B+e(E\times f) \}+c_s^2(f\times\nabla \rho_1 )]$$
\begin{eqnarray}
\label{hibus}
    \dot{\rho}_{1}= -\frac{1}{\mathcal{\mathcal{A}}}\nabla .\left( \rho_{1}\vec{v_{0}}-\rho_{0}\nabla \psi_{1}\right)~.
\end{eqnarray}
Take the time derivative of \eqref{nabol} and compare it with \eqref{hibus}, (keeping external parameters and $c_s$ fixed), to  arrive at  wave equation of  massless relativistic scalar in a curved spacetime
 \begin{equation}
     \label{effwav}
     \pa_{\mu}(f^{\mu\nu}\pa_{\nu}\psi_{1})=0 ,
 \end{equation}
%In the next parts on-wards we will use $v_{0}= u$.
$$f^{\mu\nu}= \frac{\rho_{0}}{c_s^{2}}\left(
\begin{array}{cccc}
 \mathcal{A} & v_x & v_y & v_z \\
 v_x & \frac{v_x^2-c_s^2}{\mathcal{A}} & \frac{v_x v _y}{\mathcal{A}} & \frac{v_x v_z}{\mathcal{A}} \\
 v_y & \frac{v_x v_y}{\mathcal{A}} & \frac{v_y^2-c_s^2}{\mathcal{A}} & \frac{v_y v_z}{\mathcal{A}} \\
 v_z & \frac{v_x v_z}{\mathcal{A}} & \frac{v_y v_z}{\mathcal{A}} & \frac{v_z^2-c_s^2}{\mathcal{A}} \\
\end{array}
\right)$$
%where $f^{\mu\nu}=\sqrt{-G'}G^{'\mu\nu}$
Note that $f^{\mu\nu}$ depends on  background  velocity ${v}_{0i}$ which   we write as $v_i$. Effective background metric  $g^{\mu\nu}$ is $ f^{\mu\nu}=\sqrt{-g}~g^{\mu\nu}$ with  determinant of $f^{\mu\nu}$ given by $\vert{f^{\mu\nu}}\vert={(\sqrt{-g})}^{4}\frac{1}{g}= g=-\frac{\rho_{0}^{4}}{c_{s}^{2}\mathcal{A}^{2}}~.$ The AM is constructed out of  background fluid velocity and inherits symmetries of the latter. The AM is stationary as  flow is stationary (or steady in fluid dynamics terminology). Thus cherished AM $g_{\mu\nu}$, one of our major results, in polar form is,
\begin{widetext}
  \begin{eqnarray}
g_{\mu\nu}= \frac{\rho_{0}}{\mathcal{A} c_s}\left(
\begin{array}{cccc}
 \frac{c_s^2-(v_{r}^2+v_{\theta}^{2}+v_{\phi}^{2})}{\mathcal{A} } & v_{r}& r v_{\theta} & r\sin{\theta}v_{\phi} \\
 v_{r} & -\mathcal{A} & 0 & 0 \\
 r v_{\theta} & 0 & -\mathcal{A}r^{2} & 0 \\
 r\sin{\theta}v_{\phi} & 0 & 0 & -\mathcal{A}r^{2}\sin^{2}{\theta} \\
\end{array}
\right)\nonumber
 \end{eqnarray}
 \end{widetext}
%We adopt a scenario where the background fluid flow is a planner flow so that  $v_{\theta}=0$.}}
It is important to remember that  fluid particles see the flat Minkowski metric (for fluid velocity $<<$ velocity of the electromagnetic field in vacuum) whereas acoustic fluctuations feel only the AM; some basic properties of the latter carry a legacy of the former. From the above AM it is clear \cite{vis} that regions of supersonic flow are ergo-regions where $g_{tt}$ changes sign, $g_{tt}=0 ~\rightarrow ~v_r=c_s$ corresponds to event horizon (wave-blocking zone in fluid dynamics), the boundary that null geodesics (or phonons) can not escape. In fact here ergosphere coincides with event horizon. Other notions such as trapped surface, surface gravity, etc also exist for AM \cite{vis}. Spatial positions of the analogue horizon in the fluid will appear indirectly from $c_s(r), v_r(r)$. 

 \textit{ $m_{eff}, a_{eff}$ in AM-Kerr analogy:}  For matching with Kerr,  we convert the acoustic path length dimension to  $|ds^{2}|=(\text{length})^{2}=[L]^{2}$. In GR,  metrics have dimensional parameters such as Newton's  constant $G$ and velocity of light $c$, among others. Similarly, AM can depend on $c_s$, background fluid density $\rho_0$ (both not  constant in general) etc.  Another fluid parameter is dynamic (or absolute) viscosity $\mu$ of dimension of $|\mu|=[M][L]^{-1}[T]^{-1}$ (with kinematic viscosity  being $ \mu/\rho_{0}$). A length scale $l$ ($\sim$   spatial dimension of  the fluid system) enters our acoustic model.  $\bf{k}$-dependence in $\bf{\Omega}(\bf k)$ refers to the quasi-momentum of a single band (in the crystalline solid) of the  Bloch electron, comprising the electron  fluid  in hydrodynamic limit, where the AM is constructed. For the present work, $\bf{k}$ is just a label and is treated as a constant. For a uniform $\bf{B(r)}=\bf {B}$, $\mathcal{A}$ is effectively a constant. Resulting acoustic path  has $|ds_{\text{AM}}^{2}|=(\text{length})^2$ dimension,
\begin{widetext}
	\begin{eqnarray}
		ds_{\text{AM}}^{2}=\frac{c_s l\rho_{0}}{\mu\mathcal{A}}\left[\frac{\left(c_s^{2}-v^{2}\right)}{\mathcal{A}} dt^{2} + 2v_{r} dt dr + 2rv_{\theta}dt d\theta + 2 r \sin{\theta}v_{\phi}dt d\phi  - \mathcal{A}\{dr^{2} + r^{2}d\theta^{2} + r^{2}\sin^{2}\theta d\phi^{2}\}\right]~\label{acoustic_metric_spherically_sym_with param},
	\end{eqnarray}
	\label{ac1}
%	\begin{eqnarray}
%	ds_{\text{AM}}^{2}=\frac{c_s l\rho_{0}}{\mu\mathcal{A}}\left[\frac{\left(c_s^{2}-v_{r}^{2}\right)}{\mathcal{A}} dt^{2} + 2v_{r} dt dr - \mathcal{A}\{dr^{2} + r^{2}d\theta^{2} + r^{2}\sin^{2}\theta d\phi^{2}\}\right]~.\label{acoustic_metric_spherically_sym_with param}
%	\end{eqnarray}
% \label{ac1}
\end{widetext}
where $v^{2}=v_{r}^{2}+v_{\theta}^{2}+v_{\phi}^{2}$. Now we perform a coordinate transformation 
\begin{eqnarray}
	dt\rightarrow dt+\frac{dr}{c_s} + \frac{d\theta}{\omega_s} + \frac{ d\phi}{\Omega_s},
 \\~\nonumber\omega_s = \text{angular frequency}, \\
\nonumber \Omega_s = \text{azimuthal frequency of sonic disturbance,}
  \label{generalised_transformation}
\end{eqnarray}
 on the acoustic path  to obtain 
\begin{widetext}
	\begin{eqnarray}
		 ds_{\text{AM}}^{2}=&& \frac{c_s l\rho_{0}}{\mu\mathcal{A}}\Bigg[\frac{\left(c_s^{2}-v^{2}\right)}{\mathcal{A}} dt^{2} + \left\{\frac{\left(c_s^{2}-v^{2}\right)}{\mathcal{A}c_{s}^{2}} + \frac{2v_{r}}{c_{s}^{2}} -\mathcal{A}\right \}dr^{2} + 2  \left\{\frac{\left(c_s^{2}-v^{2}\right)}{\mathcal{A}c_{s}} + v_{r} \right\}dt dr + 2 \left\{\frac{\left(c_s^{2}-v^{2}\right)}{\mathcal{A}\Omega_{s}} + r \sin{\theta} v_{\phi}\right\} dt d\phi \nonumber\\
         &&+ 2 \left\{\frac{\left(c_s^{2}-v^{2}\right)}{\mathcal{A}\omega_{s}} + r v_{\theta}\right\} dt d\theta +2 \left\{\frac{\left(c_s^{2}-v^{2}\right)}{\mathcal{A}c_{s}} + \frac{v_{r}}{\omega_{s}}+\frac{r v_{\theta}}{c_{s}} \right\}dr d\theta + 2 \left\{\frac{\left(c_s^{2}-v^{2}\right)}{\mathcal{A}\Omega_{s}\omega_{s}} + \frac{r \sin{\theta}v_{\phi}}{\omega_{s}}+\frac{r v_{\theta}}{\Omega_{s}} \right\}d\theta d\phi \nonumber\\
		&&+ 2  \left\{\frac{\left(c_s^{2}-v^{2}\right)}{\mathcal{A}c_{s}\Omega_{s} }+  \frac{v_{r}}{\Omega_{s}} +\frac{r \sin{\theta} v_{\phi}}{c_{s}}\right\}dr d \phi +\left\{ \frac{\left(c_s^{2}-v^{2}\right)}{\mathcal{A}\Omega_{s}^{2}} -\mathcal{A}r^{2}\sin^{2}{\theta} +\frac{2r\sin{\theta}}{\Omega_{s}}v_{\phi}\right\}d\phi ^{2}
  \nonumber\\
       &&+\left\{ \frac{\left(c_s^{2}-v^{2}\right)}{\mathcal{A}\omega_{s}^{2}} -\mathcal{A}r^{2} +\frac{2r}{\omega_{s}}v_{\theta}\right\}d\theta ^{2}\Bigg]~.\label{acoustic_metric_spherically_sym_with param}
	\end{eqnarray}
	%\label{ac1}
\end{widetext}
Our major observation is that  in the equatorial plane (i.e. $\theta=\pi/2$ hyper-surface) and with $v_{\theta}=0$, the acoustic path 
  \begin{widetext}
	\begin{eqnarray}
	 ds_{\text{AM}}^{2}=&& \frac{c_s l\rho_{0}}{\mu\mathcal{A}}\Bigg[\frac{\left(c_s^{2}-v^{2}\right)}{\mathcal{A}} dt^{2} + \left\{\frac{\left(c_s^{2}-v^{2}\right)}{\mathcal{A}c_{s}^{2}} + \frac{2v_{r}}{c_{s}^{2}} -\mathcal{A}\right \}dr^{2} + 2  \left\{\frac{\left(c_s^{2}-v^{2}\right)}{\mathcal{A}c_{s}} + v_{r} \right\}dt dr + 2 \left\{\frac{\left(c_s^{2}-v^{2}\right)}{\mathcal{A}\Omega_{s}} + r v_{\phi}\right\} dt d\phi\nonumber\\
		&&+ 2  \left\{\frac{\left(c_s^{2}-v^{2}\right)}{\mathcal{A}c_{s}\Omega_{s} }+  \frac{v_{r}}{\Omega_{s}} +\frac{r v_{\phi}}{c_{s}}\right\}dr d \phi +\left\{ \frac{\left(c_s^{2}-v^{2}\right)}{\mathcal{A}\Omega_{s}^{2}} -\mathcal{A}r^{2} +\frac{2r}{\Omega_{s}}v_{\phi}\right\}d\phi ^{2} \Bigg]~,\label{acoustic_metric_spherically_sym_with param}
	\end{eqnarray}
	%\label{ac1}
	\end{widetext}
   is structurally equivalent to  Kerr metric path length in Eddington-Finklestein coordinates (full expressions in Supplemental Material eqs.(5-10)).
  \begin{widetext}
  \begin{eqnarray}
	ds^{2}_{\text{Kerr}}=&& \Bigg(1-\frac{2Gm}{r c^2}\Bigg)c^2dt^{2}-\frac{4Gm}{r c}dtdr+\frac{4Gma}{r c^2} dt d\phi-\Bigg(1+\frac{2Gm}{r c^2}\Bigg)dr^{2}+2\frac{a}{c}\Bigg(1+\frac{2Gm}{r c^2}\Bigg)drd\phi\nonumber\\
	&&-\Bigg(r^{2}+\frac{a^{2}}{c^2}-\frac{2Gma^{2}}{r c^4}\Bigg) d\phi^{2}~. \label{Kerr_metric_at_pi/2}~
	\end{eqnarray}
\end{widetext}
   We  exploit the dimensional equality $ds^{2}_{\text{AM}}=ds^{2}_{\text{Kerr}}=(\text{length})^2$ to construct  effective mass and spin parameters for AM, in analogy with    mass ($m$) and angular momentum per unit mass $a=J/m$ of  Kerr black hole (with details in Supplemental Material eqs.(11-17)).
\begin{itemize}
	\item Comparison of  dimensions of $g_{tt}$ gives 
\begin{eqnarray}
 m_{\text{eff}}\equiv \frac{l^{3}\rho_{0}v^{2}}{\mathcal{A}^{2} c_s^{2}}\label{mass_eff_final}
%m_{\text{eff}}\equiv \frac{\mu l^{2} v_r^{2}}{\mathcal{A} c_s^{3}}~.\label{mass_eff_final}
%&&\boxed{m_{\text{eff}}\equiv \frac{L^{3}\rho_{0}}{\mathcal{A}^{2}}y^{2}}~~~~~~~\text{(replacing $v_r=(L/t)y$ and $t=L/c_s$)}~.
\end{eqnarray}
and
	\item comparison of dimensions of $g_{t\phi }$  gives
\begin{eqnarray}
a_{\text{eff}}\equiv \frac{l \rho_{0} c_s^{5}}{\mu \mathcal{A}^{2}\Omega_s v^{2}}~.
\end{eqnarray}
\end{itemize} 
 This constitutes another set of important results since these effective parameters are, in principle, measurable. 
			
%\begin{eqnarray}
%	ds_{\text{AM}}^{2} = && \left(\frac{c_s l \rho_{0}}{\mu\mathcal{A}^{2}} - m_{\text{eff}}\frac{c_s}{\mu l^{2}}\right)c_s^{2}dt^{2} + 2 \frac{m_{\text{eff}}}{\mu l^{2}}\left(c_s^{2} - v_{r}^{2} + c_s v_{r} \right)dt dr + 2 m_{\text{eff}} a_{\text{eff}} \left(\frac{\mu\mathcal{A}v_{r}^{2}}{c_s^{5} l \rho_{0}}\right)\left(c_s^{2} - v_{r}^{2}\right) dt d\phi\nonumber\\
%	&&+\left[\frac{c_s l \rho_{0}}{\mu\mathcal{A}}\left(1+2\frac{v_r}{c_s}-\mathcal{A}\right)-m_{\text{eff}}\frac{c_s}{\mu l^{2}} \right]dr^{2} + 2 a_{\text{eff}} \frac{v_{r}^{2}}{c_s^{2}}\left[\frac{1}{c_{s}\mathcal{A}}\left\{1-m_{\text{eff}}\frac{\mathcal{A}^{2}}{l^{3}\rho_{0}}\right\}+\frac{v_{r}}{c_s^{2}} \right]dr d\phi\nonumber\\
%	&&+\left[\frac{\mu v_{r}^{4}}{c_s^{7} l \rho_{0}}a_{\text{eff}}^{2}\left(1-m_{\text{eff}}\frac{\mathcal{A}^{2}}{l^{3}\rho_{0}}\right)  -\frac{c_s l \rho_{0}}{\mu}r^{2}\right]d\phi^{2}~.\label{final_metric_in_terms_m_a}
%\end{eqnarray}
In terms of  $m_{\text{eff}}$ and $a_{\text{eff}}$ the same metric (\ref{acoustic_metric_spherically_sym_with param}) turns out to be
\begin{widetext}
\begin{eqnarray}
ds_{\text{AM}}^{2} = && \left(\frac{c_s l \rho_{0}}{\mu\mathcal{A}^{2}} - m_{\text{eff}}\frac{c_s}{\mu l^{2}}\right)c_s^{2}dt^{2} + 2 m_\text{eff}\frac{c^{2}_{s}}{\mu l^{2}}\left(\frac{c_s^{2}}{v^2} - 1 + \frac{c_s v_{r}\mathcal{A}}{v^2} \right)dt dr + 2 m_{\text{eff}} a_{\text{eff}} \left(\frac{\mathcal{A}^{2}}{l^{3}\rho_{0}}\right)\left(1 - v^{2}+\frac{\Omega_{s}\mathcal{A}}{c_{s}^{2}}r v_{\phi}\right) dt d\phi\nonumber\\
&&+\left[\frac{c_s l \rho_{0}}{\mu\mathcal{A}^{2}}\left(1+2\frac{v_r}{c^{2}_{s}}\mathcal{A}-\mathcal{A}^{2}\right)-m_{\text{eff}}\frac{c_s}{\mu l^{2}} \right]dr^{2} + 2 a_{\text{eff}} \frac{v_{r}^{2}}{c_s^{3}}\left[1-m_{\text{eff}}\frac{\mathcal{A}^{2}}{l^{3}\rho_{0}c_{s}}+\frac{\mathcal{A}v_{r}}{c_s} +\frac{\mathcal{A}\Omega_{s}}{c^{2}_{s}}r v_\phi\right]dr d\phi\nonumber\\
&&+\left[\frac{\mu\mathcal{A}^{2} v^{4}}{c_s^{7} l \rho_{0}}a_{\text{eff}}^{2}\left(1-m_{\text{eff}}\frac{\mathcal{A}^{2}}{l^{3}\rho_{0}}\right)  -\frac{c_s l \rho_{0}}{\mu}r^{2} +2\frac{c_s l \rho_{0}}{\mu \mathcal{A} \Omega_{s}}r v_{\phi} \right]d\phi^{2}~.\label{final_metric_in_terms_m_a}
\end{eqnarray}
\end{widetext}
 Remarkably, our entirely algebraic methodology for implementing coordinate transformations
 (\ref{generalised_transformation}) and prescription of identifying fluid mass and spin parameters have resulted in an AM (\ref{final_metric_in_terms_m_a}), which can be compared term by term with  Kerr metric (see SM). Notice that $m_{eff},~a_{eff}$ in AM (\ref{final_metric_in_terms_m_a}) occupy identical positions as $m,~a$ in  Kerr metric (see SM). This provides a mathematical consistency of our framework and reveals the physics behind AM.
 \\
{\it{ Phase space probe trajectory:}} The AM  at $\theta=\pi/2$  has a timelike Killing vector $\chi^a=(1,0,0,0)$ with conserved   energy of a particle  given by $E=-\chi^ap_a=-p_t$, where $p_a=(p_t,p_r,p_\theta,p_\phi)$. Using the AM in   the dispersion relation $g^{ab}p_ap_b=-M^2$ for a  particle of  mass $M$ in AM  (with $p_{\theta}=0$)), particle energy $E$ is obtained in terms of the other momentum components, as the positive energy root.  Hamilton's equations of motion are 
\begin{equation}
\dot{r}=\frac{\partial E}{\partial p_r},~
\dot{p_r} = -\frac{\partial E}{\partial r},~
\dot{\phi} = \frac{\partial E}{\partial p_\phi},~
\dot{p_\phi} = -\frac{\partial E}{\partial\phi}.\label{a6}
\end{equation}
Let us consider a particular background fluid profile known as ``draining bathtub" flow (for details see \cite{vis})
\begin{eqnarray}
	{\bf{v}}= \frac{A\hat{r}+B\hat{\phi}}{r}\label{background flow}~
\end{eqnarray}
with constant $A,B$. In this idealized model, background fluid flow is planar until it reaches  a linear sink along  perpendicular direction. The background fluid density $\rho$ is taken to be constant throughout the flow. Furthermore, for the  barotropic fluid considered here, (\ref{beul}) and specific enthalpy ($h$) indicate that  background pressure $P$ and  speed of  sonic disturbance $c_{s}$ are also constant. The equation of continuity (\ref{bcont}), in cylindrical coordinates with sink along $z$-direction, reduces to 
\begin{equation}
\frac{1}{r}(\frac{\partial}{\partial r}(rv_r) +\frac{\partial v_{\phi}}{\partial \phi} +\frac{\partial}{\partial z}(rv_z) )=0 ,
    \label{ceq}
\end{equation}
and clearly the profile (\ref{background flow}) (with $v_z=0$ on the plane just away from the sink and planar distance $r$ measured from $z$-axis) is a solution of (\ref{ceq}) (for more details, see see  section 2.4.3 of \cite{vis}.

In this model the acoustic ergosphere and event horizon form at
	\begin{equation}
		r_{\text{ergosphere}} = \frac{\sqrt{A^{2}+B^{2}}}{c_{s}}, ~~ r_{\text{horizon}}=\frac{|A|}{c_{s}} . \label{horizon position}
	\end{equation} 
However, one important thing needs be to mentioned here is that the distinction between ergosphere and the acoustic horizon is critical for this model \cite{vis}. Therefore, keeping that in mind, we proceed here to solve these coupled differential equations numerically (Eqns. (\ref{a6})).

 After the numerical solutions we have plotted the phase-space plot between the radial coordinate $(r)$ and the corresponding radial momentum of the particle $(p_r)$. Depending on the sign of $A$ (+ and -) we have plotted two cases. The values of the other parameters are as follows, $\mathcal{A}=5$, $c_{s}=100$, $\Omega_{s}=1.0$, $\Gamma=\frac{c_{s}l\rho_{0}}{\mu\mathcal{A}}=100$. In the first figure, i.e. Fig. (\ref{fig:massless}) the amplitudes of the velocity components (i.e. $v_r$ and $v_\phi$) are respectively $A=B=100000$. In Fig. (\ref{fig:massless}), we can see, that the phase-space trajectory of the particle starts with some lower momentum value for a larger value of $r$, but as $r$ decreases, the corresponding radial momentum value ($p_{r}$) increases and at $r=1000$ the momentum reaches its maximum value. This nature of the graph depicts that as the particle moves near to the $r=1000$ the particle experiences a "sudden change" in its trajectory. Moreover, according to the "draining bathtub" model the acoustic horizon should appear at $r=1000$ (see Eqn. (\ref{horizon position})) which is exactly happening in our case based on our specified parameter values. Consequently, from this occurrence, we can identify the position of the horizon, which exactly matches the theoretical value of the horizon, i.e. at $|A|/c_s=1000$. In this context, it is worth mentioning that in some near-horizon context \cite{Dalui2022,Dalui2019,Dalui:2020qpt} it has been shown that in the near-horizon region, a particle experiences this kind of "sudden change" or "instability" in its phase-space trajectory.

 Similarly, in Fig. (\ref{fig:ingoing}) we have chosen $A=-100000$ and $B=500$ keeping the other parameters the same and we found that the radial momentum value of the massless particle does not change much until reaching $r\simeq2000$. After $r\simeq1000$ the momentum value falls abruptly which suggests that the particle is sucked inside the horizon which is situated at $r=1000$. This characteristic is exactly similar to an ingoing massless particle in the near-horizon region of a SSS BH (see Eq. (26) and page no. 8 of \cite{Dalui:2020qpt}) and a Kerr BH (see page no. 6 of \cite{Dalui2022}).

 %Moreover, this particular nature of the graph (starting from the horizon) mimics the phase-space diagram of a $XP$ like Hamiltonian \cite{Berry}. Furthermore, we know people have already shown \cite{Dalui2022,Dalui2019} that in the near-horizon region of a Kerr BH or any static spherically symmetric BH in EF coordinates, the Hamiltonian of a massless particle becomes exactly like $XP$ like which suggests the unstable region in the near-region. Not only that, the near horizon instability comes in a disguise of the Hamiltonian of inverse harmonic oscillator (IHO) also \cite{Hashimoto:2016dfz,Hegde:2018xub} and as we know $XP$ is also transformed into the Hamiltonian of IHO in some canonically rotated coordinate systems \cite{Dalui2019}. Therefore, everything falls into this figure, we not only determine the position of the horizon but also capture the exact characteristics of a massless particle in the near-horizon region. Hence, we can conclusively say that our model exactly mimics the characteristics of a Kerr black hole.}
 
%\textcolor{red}{For the massive case also, we can see that the particle experiences the instability near to the acoustic horizon ($r\simeq 1000$). In some references it has been shown that this instability sometime leads to chaotic phenomena for some particular parameter ranges \cite{Dalui:2018qqv}. However, in the present scenario we are not going to discuss that and we leave that for future discussions.}   
\begin{widetext}

\begin{figure}[H]
	\centering
	\includegraphics[scale=0.19, angle=0]{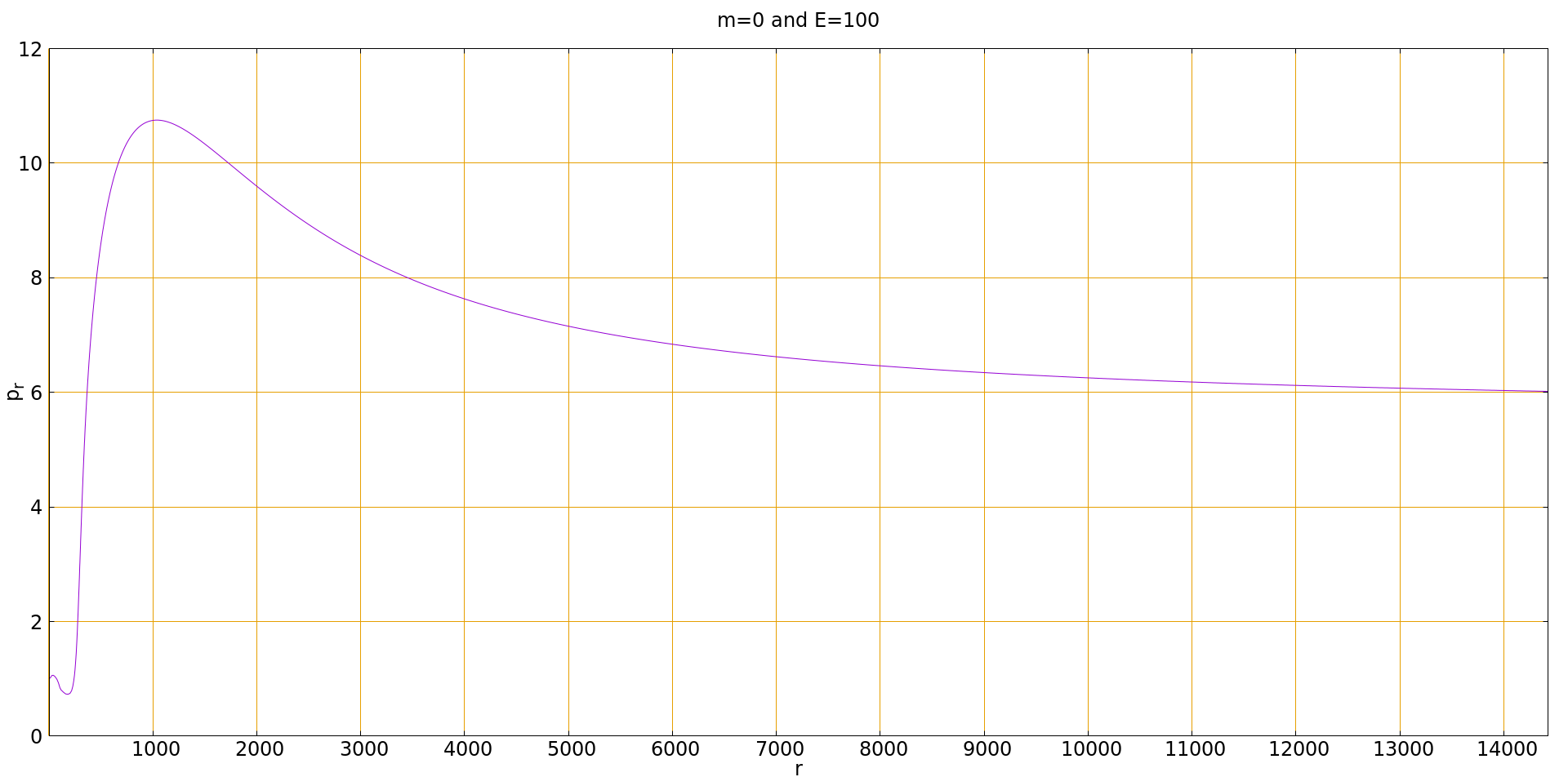}
	\caption{Phase-space diagram for  a massless particle (for $A=B=1000000$). From the figure we can see as the particle moves near to $r=1000$ its radial momentum $p_{r}$ increases exponentially reaching its peak as $r$ attains the value of 1000. This occurrence enables us to pinpoint the horizon's position, which precisely corresponds to the theoretical expectation at $|A|/c_s=1000$. }
	\label{fig:massless}
\end{figure}

\begin{figure}[H]
	\centering
	\includegraphics[scale=0.2, angle=0]{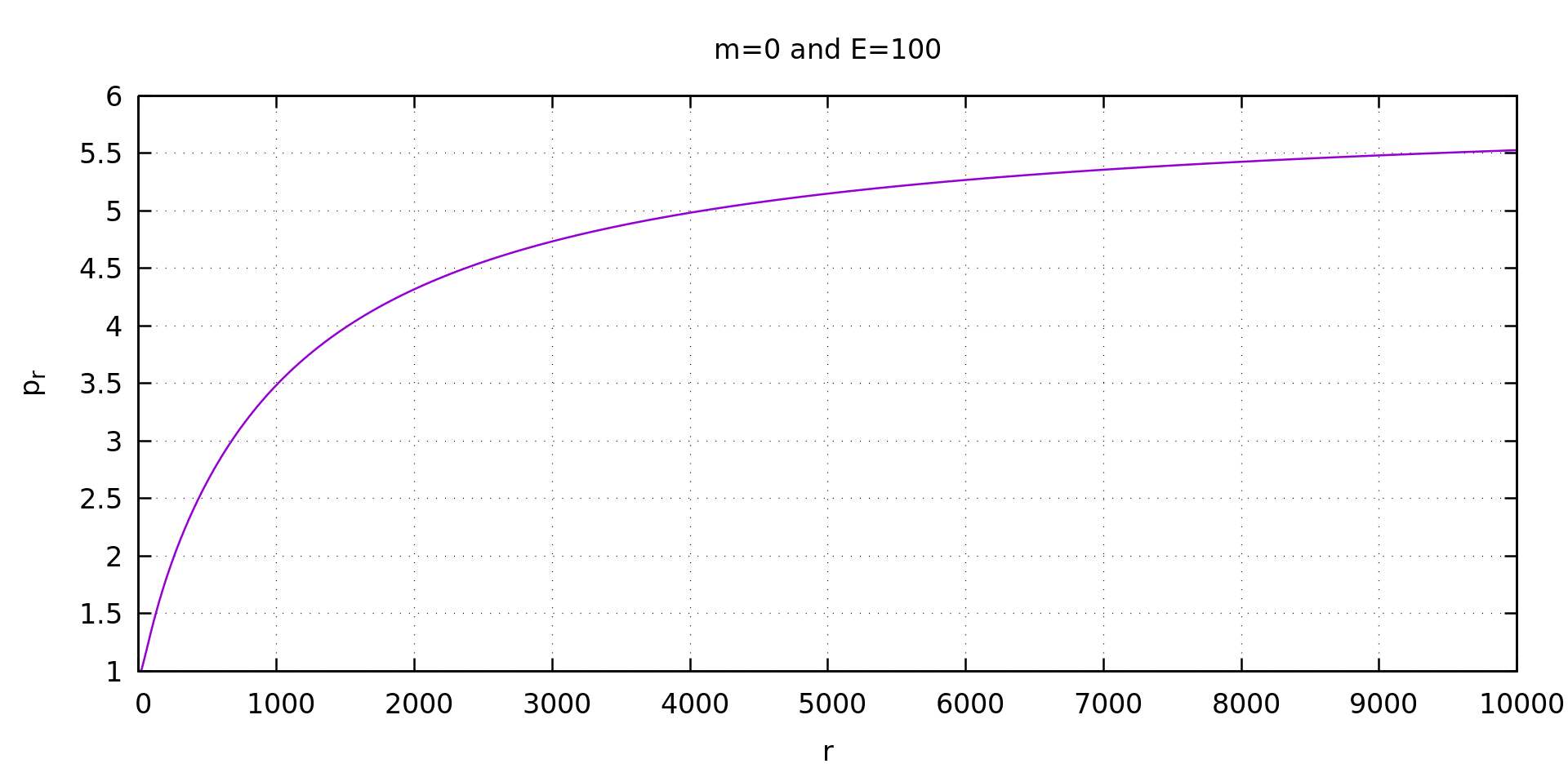}
	\caption{Similarly, for $A=-100000$ and $B=500$ we plot again plot the phase-space trajectory of the massless particle remaining other parameter values the same. We see that until around $r\simeq2000$ the radial momentum of the particle does not change much but near to $r\simeq1000$ the momentum of the particle suddenly falls down which suggests that the massless particle falls inside the horizon.}
	\label{fig:ingoing}
\end{figure}    
\end{widetext}
In a nutshell, we can say that with the radial dependent background flow of fluid, we have constructed an analogue metric which can mimic the exact structure of Kerr BH and by studying the particle dynamics in this background we can pinpoint the exact location of the horizon for particular values.
 Future works will involve a more rigorous analysis with the full anomalous fluid dynamics and an arbitrary fluid flow. Attempts of laboratory demonstrations of this new analogue Black Hole model will be worthwhile.

\begin{widetext}
	\section*{Supplemental material}
	 \noindent
	Full form of the extended fluid model with Berry curvature corrections, derived by two of the present authors in \cite{mitra2021}, in terms of continuity and Euler equations are
	\setcounter{equation}{0}
	\begin{equation}
		\label{con}
		\dot{\rho}+{\bf{\nabla}}\cdot{\bf J}=0
	\end{equation}
	\begin{equation}
		\label{J}
		{\bf J}=\left(\frac{\rho{\bf{v}}}{\mathcal{A}}\right)+ e \rho({\bf{\mathcal{F}}}\cdot{\bf{v}}){\bf{B}}+\mathcal{F}\times \nabla P ,
	\end{equation}
	
	\begin{eqnarray}
		\dot{{\bf{v}}}+ \frac{({\bf{v}}\cdot{\bf{\nabla}}){\bf{v}}}{\mathcal{A}} = && -\frac{{\bf{\nabla}}P}{\rho\mathcal{A}} +e\frac{\rho {\bf{v}}\times{\bf{B}}}{\rho\mathcal{A}}-e \frac{{\bf{B}\cdot{\bf{\nabla}}P}}{\rho}{\bf{\mathcal{F}}}-e({\bf{v}}\cdot{\bf{\mathcal{F}}})
		({\bf{B}}\cdot{\bf{\nabla}}){\bf{v}}+\Bigg[\left(\frac{\nabla P}{\rho}\times \mathcal{F}\right)\cdot\nabla
		-\frac{1}{\rho}\nabla v^2 \cdot\{\nabla \times  {\bf({\mathcal{F}\rho)}}\}\nonumber\\
		&&+ 2 v^2 \left({\bf{\mathcal{F}}}\times\frac{\nabla \rho}{\rho}\right)\cdot\nabla - {\bf{\mathcal{F}}}\cdot \left(\frac{\nabla \rho}{\rho}\times \nabla v^2\right)\Bigg]{\bf{v}}, \label{eulber}
	\end{eqnarray}	
	%\end{widetext}	
	\begin{equation}
		\mathcal{F}_{i}({\bf x,\bf k})=\frac{\Omega_{i}}{1+e{\bf{B}}({\bf x})\cdot{\bf{\Omega}(\bf k)} }, 
		\mathcal{A}({\bf x,\bf k})=1+e{\bf{B}}({\bf x})\cdot{\bf{\Omega}(\bf k)}~.
	\end{equation}
	%$$\mathcal{F}_{i}({\bf x,\bf k})=\frac{\Omega_{i}}{1+e{\bf{B}}({\bf x})\cdot{\bf{\Omega}(\bf k)} }~~~\text{and}~~	\mathcal{A}({\bf x,\bf k})=1+e{\bf{B}}({\bf x})\cdot{\bf{\Omega}(\bf k)}.$$
	The reduced form used in the paper is obtained for small ${\bf\Omega(k)} $ such that we neglect $\mathcal{F}$  multiplied by any first order term. Furthermore as shown in  \cite{mitra2021}, this fluid model has an Adler-Bell-Jackiw like divergence anomaly, proportional to external electric field. But in the present work We will not consider any anomaly effect since for simplicity, we do not have any external electric field in the present model.
	
	In order to facilitate a meaningful identification of dimensional parameters of our acoustic metric with a metric in GR, we follow two steps. In the first step, we convert the acoustic path length dimension to  $|ds^{2}|=(\text{length})^{2}=[L]^{2}$. In GR, the metrics can comprise dimensional parameters such as Newtonian gravitational constant $G$ and velocity of light $c$, among others. In a similar way, the Acoustic metric can depend on $c_s$ (not a constant in general), background fluid density $\rho_0$ (not a constant in general) etc.  This scheme requires us to introduce another conventional fluid parameter known as the dynamic (or absolute) viscosity $\mu$ of dimension of $|\mu|=[M][L]^{-1}[T]^{-1}$ (with the kinematic viscosity  being $ \mu/\rho_{0}$). We also introduce a length scale $l$ (which can be the spatial dimension of the fluid) into our acoustic system. This yields the acoustic metric    
	
	\begin{eqnarray}
		ds_{\text{AM}}^{2}=\frac{c_s l\rho_{0}}{\mu\mathcal{A}}\left[\frac{\left(c_s^{2}-v^{2}\right)}{\mathcal{A}} dt^{2} + 2v_{r} dt dr + 2rv_{\theta}dt d\theta + 2 r \sin{\theta}v_{\phi}dt d\phi  - \mathcal{A}\{dr^{2} + r^{2}d\theta^{2} + r^{2}\sin^{2}\theta d\phi^{2}\}\right]~\label{acoustic_metric_spherically_sym_with param}
	\end{eqnarray}

	%\begin{widetext}
	%	\begin{eqnarray}
		%	ds_{\text{AM}}^{2}=\frac{c_s l\rho_{0}}{\mu\mathcal{A}}\left[\frac{\left(c_s^{2}-v_{r}^{2}\right)}{\mathcal{A}} dt^{2} + 2v_{r} dt dr - \mathcal{A}\{dr^{2} + r^{2}d\theta^{2} + r^{2}\sin^{2}\theta d\phi^{2}\}\right]~.\label{acoustic_metric_spherically_sym_with param}
		%	\end{eqnarray}
	%\\\end{widetext}
such that $|ds_{\text{AM}}^{2}|=(\text{length})^2$.

In the second step,  we perform a coordinate transformation 
%\begin{widetext}
\begin{eqnarray}
	dt\rightarrow dt+\frac{dr}{c_s} + \frac{d\theta}{\omega_s} + \frac{ d\phi}{\Omega_s}.\label{generalised_transformation}
\end{eqnarray}
%\end{widetext}
where $c_s$, $\omega_s$ and $\Omega_s$ are the sound velocity, the angular frequency and the azimuthal frequency of the sonic disturbance. The Acoustic metric is transformed to 

%\begin{widetext}
\begin{eqnarray}
	ds_{\text{AM}}^{2}=&& \frac{c_s l\rho_{0}}{\mu\mathcal{A}}\Bigg[\frac{\left(c_s^{2}-v^{2}\right)}{\mathcal{A}} dt^{2} + \left\{\frac{\left(c_s^{2}-v^{2}\right)}{\mathcal{A}c_{s}^{2}} + \frac{2v_{r}}{c_{s}^{2}} -\mathcal{A}\right \}dr^{2} + 2  \left\{\frac{\left(c_s^{2}-v^{2}\right)}{\mathcal{A}c_{s}} + v_{r} \right\}dt dr + 2 \left\{\frac{\left(c_s^{2}-v^{2}\right)}{\mathcal{A}\Omega_{s}} + r \sin{\theta} v_{\phi}\right\} dt d\phi \nonumber\\
	&&+ 2 \left\{\frac{\left(c_s^{2}-v^{2}\right)}{\mathcal{A}\omega_{s}} + r v_{\theta}\right\} dt d\theta +2 \left\{\frac{\left(c_s^{2}-v^{2}\right)}{\mathcal{A}c_{s}} + \frac{v_{r}}{\omega_{s}}+\frac{r v_{\theta}}{c_{s}} \right\}dr d\theta + 2 \left\{\frac{\left(c_s^{2}-v^{2}\right)}{\mathcal{A}\Omega_{s}\omega_{s}} + \frac{r \sin{\theta}v_{\phi}}{\omega_{s}}+\frac{r v_{\theta}}{\Omega_{s}} \right\}d\theta d\phi \nonumber\\
	&&+ 2  \left\{\frac{\left(c_s^{2}-v^{2}\right)}{\mathcal{A}c_{s}\Omega_{s} }+  \frac{v_{r}}{\Omega_{s}} +\frac{r \sin{\theta} v_{\phi}}{c_{s}}\right\}dr d \phi +\left\{ \frac{\left(c_s^{2}-v^{2}\right)}{\mathcal{A}\Omega_{s}^{2}} -\mathcal{A}r^{2}\sin^{2}{\theta} +\frac{2r\sin{\theta}}{\Omega_{s}}v_{\phi}\right\}d\phi ^{2} \nonumber\\
	&&+\left\{ \frac{\left(c_s^{2}-v^{2}\right)}{\mathcal{A}\omega_{s}^{2}} -\mathcal{A}r^{2} +\frac{2r}{\omega_{s}}v_{\theta}\right\}d\theta ^{2}\Bigg]~.\label{acoustic_metric_spherically_sym_with param}
\end{eqnarray}
%\label{ac1}
%\end{widetext}
\begin{comment}

\begin{widetext}
	\begin{eqnarray}
		ds^{2}_{\text{AM}}=&&\frac{c_s l\rho_{0}}{\mu\mathcal{A}}\Bigg[\frac{\left(c_s^{2}-v_{r}^{2}\right)}{\mathcal{A}} dt^{2}   +  2\left\{\frac{(c_s^{2}-v_{r}^{2})}{c_s\mathcal{A}} + v_{r}\right\} dtdr +   2\left\{ \frac{(c_s^{2}-v_{r}^{2})}{\omega_s\mathcal{A}}\right\} dtd\theta +  2\left\{ \frac{(c_s^{2}-v_{r}^{2})}{\Omega_s\mathcal{A}}\right\} dtd\phi \nonumber\\
		&& + \left\{\frac{\left(c_s^{2}-v_{r}^{2}\right)}{c_s^{2}\mathcal{A}} + 2\frac{v_{r}}{c_s} - \mathcal{A}  \right\} dr^{2}  + \frac{2}{\omega_s}\left\{\frac{\left(c_s^{2}-v_{r}^{2}\right)}{c_s \mathcal{A}} + v_{r}  \right\}drd\theta  +  \frac{2}{\Omega_s}\left\{\frac{\left(c_s^{2}-v_{r}^{2}\right)}{c_s \mathcal{A}} +  v_{r}  \right\}drd\phi\nonumber\\
		&& +  \left\{ \frac{(c_s^{2}-v_{r}^{2})}{\omega_s^{2}\mathcal{A}} -\mathcal{A} r^{2} \right\}d\theta^{2} + 
		2\left\{ \frac{(c_s^{2}-v_{r}^{2})} {\omega_s\Omega_s\mathcal{A}}\right\}d\theta d\phi + \left\{ \frac{(c_s^{2}-v_{r}^{2})}{\Omega_s^{2}\mathcal{A}} - \mathcal{A}r^{2}\sin^{2}\theta \right\}d\phi^{2}\Bigg]~.\label{acoustic_metric_with_ac_param}
	\end{eqnarray} 
\end{widetext}
\end{comment}
We now perform the final task: the construction  of analogue fluid parameters to be identified with a suitable GR metric. We start by writing down the Kerr metric that represents a stationary, axisymmetric black hole  in Eddington-Finkelstein (EF) coordinates (with the metric signature (+,-,-,-)) \cite{Dalui2022}
%\begin{widetext}
\begin{eqnarray}
	ds^{2}_{\text{Kerr}}=&&\Bigg(1-\frac{2Gmr}{c^2\Sigma^{2}}\Bigg)c^2dt^{2}-\frac{4Gmr}{c\Sigma^{2}}dtdr+\frac{4Gmra}{c^2\Sigma^{2}}\sin ^{2} \theta dt d\phi-\Bigg(1+\frac{2Gmr}{c^2\Sigma^{2}}\Bigg)dr^{2}+2\frac{a}{c}\sin^{2}\theta\Bigg(1+\frac{2Gmr}{c^2\Sigma^{2}}\Bigg)drd\phi\nonumber\\
	&&-\Sigma^{2}d\theta^{2}-\Bigg(r^{2}+\frac{a^{2}}{c^2}+\frac{2Gmra^{2}\sin^{2} \theta}{c^4\Sigma^{2}}\Bigg)\sin^{2}\theta d\phi^{2}~. \label{Kerr_metric}
\end{eqnarray}
%\end{widetext}
where $m$ is the mass of the Kerr black hole, $a=J/m$ is the angular momentum per unit mass and
$\Sigma^{2}=r^{2}+a^{2}\cos^{2}\theta$. 

Now comes the  interesting part. The Acoustic (\ref{acoustic_metric_spherically_sym_with param}) and Kerr (\ref{Kerr_metric}) metrics are structurally similar on the $\theta=\pi/2$ hyper-surface, having the explicit forms shown below:
%\begin{widetext}
\begin{eqnarray}
	ds_{\text{AM}}^{2}=&& \frac{c_s l\rho_{0}}{\mu\mathcal{A}}\Bigg[\frac{\left(c_s^{2}-v^{2}\right)}{\mathcal{A}} dt^{2} + \left\{\frac{\left(c_s^{2}-v^{2}\right)}{\mathcal{A}c_{s}^{2}} + \frac{2v_{r}}{c_{s}^{2}} -\mathcal{A}\right \}dr^{2} + 2  \left\{\frac{\left(c_s^{2}-v^{2}\right)}{\mathcal{A}c_{s}} + v_{r} \right\}dt dr + 2 \left\{\frac{\left(c_s^{2}-v^{2}\right)}{\mathcal{A}\Omega_{s}} + r v_{\phi}\right\} dt d\phi\nonumber\\
	&&+ 2  \left\{\frac{\left(c_s^{2}-v^{2}\right)}{\mathcal{A}c_{s}\Omega_{s} }+  \frac{v_{r}}{\Omega_{s}} +\frac{r v_{\phi}}{c_{s}}\right\}dr d \phi +\left\{ \frac{\left(c_s^{2}-v^{2}\right)}{\mathcal{A}\Omega_{s}^{2}} -\mathcal{A}r^{2} +\frac{2r}{\Omega_{s}}v_{\phi}\right\}d\phi ^{2} \Bigg]~,\label{acoustic_metric_spherically_sym_with param}
\end{eqnarray}
%\label{ac1}
%\end{widetext}
\begin{comment}

\begin{widetext}
	\begin{eqnarray}
		ds^{2}_{\text{AM}}=&&\frac{c_s l\rho_{0}}{\mu\mathcal{A}}\Bigg[\frac{\left(c_s^{2}-v_{r}^{2}\right)}{\mathcal{A}} dt^{2}   +  2\left\{\frac{(c_s^{2}-v_{r}^{2})}{c_s\mathcal{A}} + v_{r}\right\} dtdr +  2\left\{ \frac{(c_s^{2}-v_{r}^{2})}{\Omega_s\mathcal{A}}\right\} dtd\phi + \left\{\frac{\left(c_s^{2}-v_{r}^{2}\right)}{c_s^{2}\mathcal{A}} + 2\frac{v_{r}}{c_s} - \mathcal{A}  \right\} dr^{2} \nonumber\\
		&&    +  \frac{2}{\Omega_s}\left\{\frac{\left(c_s^{2}-v_{r}^{2}\right)}{c_s \mathcal{A}} +  v_{r}  \right\}drd\phi + \left\{ \frac{(c_s^{2}-v_{r}^{2})}{\Omega_s^{2}\mathcal{A}} - \mathcal{A}r^{2} \right\}d\phi^{2}\Bigg]~,\label{acoustic_metric_with_ac_param_at_pi/2}
	\end{eqnarray} 
\end{widetext}
\end{comment}
%\begin{widetext}
\begin{eqnarray}
	ds^{2}_{\text{Kerr}}=&&\Bigg(1-\frac{2Gm}{r c^2}\Bigg)c^2dt^{2}-\frac{4Gm}{r c}dtdr+\frac{4Gma}{r c^2} dt d\phi-\Bigg(1+\frac{2Gm}{r c^2}\Bigg)dr^{2}+2\frac{a}{c}\Bigg(1+\frac{2Gm}{r c^2}\Bigg)drd\phi\nonumber\\
	&&-\Bigg(r^{2}+\frac{a^{2}}{c^2}-\frac{2Gma^{2}}{r c^4}\Bigg) d\phi^{2}~. \label{Kerr_metric_at_pi/2}~
\end{eqnarray}
%\end{widetext}
We can now exploit the dimensional equality $ds^{2}_{\text{AM}}=ds^{2}_{\text{Kerr}}=(\text{length})^2$ to construct the analogue mass and spin parameters for Acoustic metric, using only the fluid parameters introduced here.
\begin{itemize}
\item Comparing the dimensions of $g_{tt}$  we get
\begin{eqnarray}
	\text{dim}\Bigg\vert\Bigg(1-\frac{2Gm}{r c^2}\Bigg)c^2\Bigg\vert =\text{dim}\Bigg\vert \frac{c_s l\rho_{0}}{\mu\mathcal{A}^{2}}\Bigg[\left(1-\frac{v^{2}}{c_s^{2}}\right)c_s^{2}\Bigg]\Bigg\vert~.\label{dt2_term}
\end{eqnarray}
%As we are going to use the dimensional analysis and we can see that the conformal factor $\frac{c_s l\rho_{0}}{\mu\mathcal{A}}$ itself is a dimensionless quantity, so we can write 
%\begin{eqnarray}
%\Bigg(1-\frac{2Gm}{r c^2}\Bigg)c^2\equiv \Bigg[\left(1-\frac{v_{r}^{2}}{c_s^{2}}\right)\frac{c_s^{2}}{\mathcal{A}}\Bigg]~.
%\end{eqnarray} 	
so that 
%\begin{eqnarray}
%	m_{\text{eff}}\equiv \left(\frac{r c^{2}}{G} \right) \left(\frac{v_{r}^{2}}{\mathcal{A}c_{s}^{2}}\right)~.\label{mass_eff}
%\end{eqnarray}
%\begin{comment}
\begin{eqnarray}
	\text{dim} |m_{\text{eff}}|\equiv \text{dim} \Bigg\vert\left(\frac{r c^{2}}{G} \right) \left( \frac{c_s l \rho_{0} v^{2}}{\mu \mathcal{A}^{2} c_{s}^{2}}\right)\Bigg\vert~.\label{mass_eff}
\end{eqnarray}
%\end{comment}
Dimension of $r c^{2}/G$ is $[L][L^{2} T^{-2}]/[L^{3} T^{-2} M^{-1}]$, so that
\begin{equation}
	\text{dim}\Bigg\vert\frac{rc^{2}}{G}\Bigg\vert 
	=\text{dim}\Bigg\vert \frac{\mu l^{2}}{c_s}\Bigg\vert~.
	\label{gtt}
\end{equation}
Therefore, from (\ref{mass_eff}) we can write down the effective mass parameter of the acoustic metric in terms of the fluid parameters,
\begin{eqnarray}
	m_{\text{eff}}\equiv \frac{l^{3}\rho_{0}v^{2}}{\mathcal{A}^{2} c_s^{2}}\label{mass_eff_final}
	%m_{\text{eff}}\equiv \frac{\mu l^{2} v_r^{2}}{\mathcal{A} c_s^{3}}~.\label{mass_eff_final}
	%&&\boxed{m_{\text{eff}}\equiv \frac{L^{3}\rho_{0}}{\mathcal{A}^{2}}y^{2}}~~~~~~~\text{(replacing $v_r=(L/t)y$ and $t=L/c_s$)} .
\end{eqnarray}
\item Comparing the dimensions of $g_{t\phi }$  we get 
\begin{eqnarray}
	\text{dim}\Bigg\vert	\frac{4Gma}{r c^2} \Bigg\vert =  \text{dim}\Bigg\vert \frac{c_s l\rho_{0}}{\mu\mathcal{A}^{2}}   \left\{ 2 \frac{(c_s^{2}-v^{2})}{\Omega_s\mathcal{A}}\right\}\Bigg\vert~.\label{dtdphi_term}
\end{eqnarray}
Thus the effective rotation parameter can be written as, (ignoring the 1/2 factor as this is dimensionless),
\begin{eqnarray}
	\text{dim} |a_{\text{eff}}|\equiv \text{dim} \Bigg\vert\left(\frac{r c^{2}}{G m}\right)\left\{ \frac{c_s l\rho_{0}}{\mu\mathcal{A}^{2}} \right\}  \left[ \frac{(c_s^{2}-v^{2})}{\Omega_s\mathcal{A}}\right]\Bigg\vert~.
\end{eqnarray}	
Therefore, replacing $m$ by $m_{eff}$ from  (\ref{mass_eff_final}) the final expression of the analogue rotation parameter for  fluid is
\begin{comment}
	\begin{eqnarray}
		a_{\text{eff}}\equiv \frac{\mu \mathcal{A} c_s^{3}}{l\rho_{0}\Omega_s v_{r}^{2}}~.
	\end{eqnarray}
	
	\begin{eqnarray}
		a_{\text{eff}}\equiv \frac{c_s^{4}}{\Omega_s v_{r}^{2}}~.
	\end{eqnarray}	
\end{comment}
\begin{eqnarray}
	a_{\text{eff}}\equiv \frac{l \rho_{0} c_s^{5}}{\mu \mathcal{A}^{2}\Omega_s v^{2}}~.
\end{eqnarray}
\end{itemize} 
%In general, the effective parameters can depend on position but as special cases, they can be treated as constants.
\end{widetext}

\vskip 1cm
%\section{References:}

\begingroup
\renewcommand{\section}[2]{}% Remove the section heading

\end{document}